\documentclass{PoS}

\title{A Cornucopia of Experiment at RHIC}

\ShortTitle{A Cornucopia of Experiment at RHIC}

\author{\speaker{Robert D. Pisarski}\thanks{I thank P. Sorenson, M. Tannenbaum, and G. Torrieri for their comments.  This work was supported under U.S. Department of Energy grant \#DE-AC02-98CH10886.  I also thank
the Alexander von Humboldt Foundation for their support.}\\
        Dept. of Physics, Brookhaven National Laboratory, Upton, NY 11973 USA\\
        E-mail: \email{pisarski@bnl.gov}}

\abstract{I outline experimental results on heavy
ion collisions at the Relativistic Heavy Ion Collider for a non-technical
audience.  This includes: elliptic flow and nearly ideal hydrodynamics; the
suppression of hard particles and the ratio $R_{\rm AA}$;
and electromagnetic signals, including dileptons and direct photons.
Especially puzzling is why
the behavior of heavy (charm) quarks appears
to be so similar to that of light quarks.}

\FullConference{The XXVI International Symposium on Lattice Field Theory\\
		 July 14-19 2008\\
		 Williamsburg, Virginia, USA}

\begin{document}

\section{Introduction}

The study of the
collisions of heavy nuclei at high energies has a simple motivation:
heavy nuclei are big.  Either gold or lead nuclei
have $A \sim 200$ nucleons, where $A$ is the atomic number.
The diameter of such a nucleus is $A^{1/3} \sim 6$
larger than that of a proton; the transverse area,
$A^{2/3} \sim 34$ times larger.  At
high energies, one might hope to study the phase transition(s) 
possible in QCD, to a deconfined, chirally
symmetric state of matter, the Quark Gluon Plasma (QGP).  
For big nuclei, one might close to a system in thermal equilibrium.

As in other areas of hadronic physics \cite{bjorken1}, 
an essential insight was due to Bjorken \cite{bjorken2}, who
suggested that it would be useful going to energies
where a plateau in rapidity first emerges.  This is the reason
why the maximum energy of the Relativistic Heavy Ion Collider (RHIC)
at Brookhaven National Laboratory (BNL) was chosen to be what it is,
as results from the ISR at CERN had shown that 
in proton-proton ($pp$) collisions, a central plateau should emerge by then.
He also suggested that the study of hard
particles, with momentum much larger than the temperature, would be
especially useful.  

The results from RHIC have demonstrated, far beyond expectation,
signs for a novel phase at high energy density
\cite{phenix,star,brahms,phobos}.  Whatever has been 
created in the collisions of two nuclei
($AA$ collisions), it is --- {\it experimentally} ---
very unlike what happens in $pp$ collisions.
Indeed, there has
been such a profusion of experimental results that one may speak of
a ``cornucopia'' of data, whence my title.  In this talk I try to
give a brief overview of the experimental situation to date.
I generally assume that the reader is familiar with concepts
from high energy physics, such as rapidity and the like, but is unfamiliar
with the concepts essential for understanding $AA$ collisions,
such as the difference between central and peripheral collisions.
For reasons of space, I could not discuss many interesting (and still
puzzling!) features of the data.  I have tried to show the standard
plots which have come to define the field since RHIC turned on in 2000.

While I concentrate on results from RHIC, there is
continuity of results from the SPS at CERN, to RHIC.  This includes
those for $J/\Psi$ suppression and
the dilepton enhancement at low invariant mass.  What is gained
by the higher energies at RHIC is that the production of hard particles
is much more common.  That, and having a dedicated machine and experiments
which are able to intensively study the physics.

Results from RHIC will continue with an increase in the luminosity by an
order of magnitude, and
upgrades to the PHENIX \cite{phenix} and STAR \cite{star} detectors.
In the next year or so there will also be
results for heavy ions at the Large Hadron Collider (LHC) at CERN, which
will probe a significantly higher regime in energy.  As I mention later,
the physics for $AA$ collisions at the LHC {\it might} be very different 
from that at RHIC.  

While I suggest that RHIC is manifestly a triumph for experiment,
the theoretical situation is still most unsettled \cite{reviews}, 
and so I only discuss
it in passing.  In some ways, the results are analogous to those
for high-$T_c$ superconductivity, where experiment also
continues to confound theory.  I do think that with the intense study
possible at RHIC and the LHC, 
that a common theoretical basis will eventually emerge.  In all of this,
results from numerical simulations on the lattice form
the absolute bedrock upon which our understanding is based \cite{lattice}.

\section{Basics of $AA$ collisions}

At RHIC one can study $pp$, $AA$, and $dA$
collisions, where the latter are the collisions of deuterons with nuclei.
(Deuterons are used instead of protons because the charge/mass ratio 
is closer to that of a large nucleus.)  For $pp$ and $AA$ collisions, the basic
variable is the energy per nucleon, $\sqrt{s}/A$.  At the AGS at Brookhaven,
this goes up to $5$~GeV; up to $17$~GeV at the SPS at CERN; and from
$20$ to $200$~GeV at RHIC.  When I quote results from RHIC, typically
I shall quote values from the highest energies, $200$~GeV.
To date, there do not appear to be dramatic differences in
going from the lowest, to the highest energies at RHIC.  This will
be clarified in the coming years with low energy runs at RHIC down to
$\sim 5$~GeV.

At the highest energies at RHIC 
there is no nuclear stopping: the incident nucleons
go down the beam pipe.  Instead of the momentum along the beam, $p_z$, it is
better to use the rapidity, $y = 1/2 \log((E+p_z)/(E-p_z))$, which
transforms additively under Lorentz boosts along the beam.  Thus one considers
the distribution of particles at a given rapidity,
$y$, versus transverse momentum, $p_t$.  
Typically I concentrate on results at zero rapidity, $p_z = y = 0$.

The AGS and SPS are fixed target machines; RHIC and LHC are colliders.
Fixed target machines allow for much higher luminosities, but it is then
difficult to study zero rapidity, since that it somewhere in the forward
direction.  For colliders, zero rapidity is at 90$^o$ to the beam, 
facilitating detector construction.  
The central plateau, being essentially free of the incident baryons, is
the most natural place to produce a system at nonzero temperature, and
(almost) zero quark chemical potential \cite{bjorken2}

\begin{figure}
\begin{center}
\includegraphics[width=0.5\textwidth]{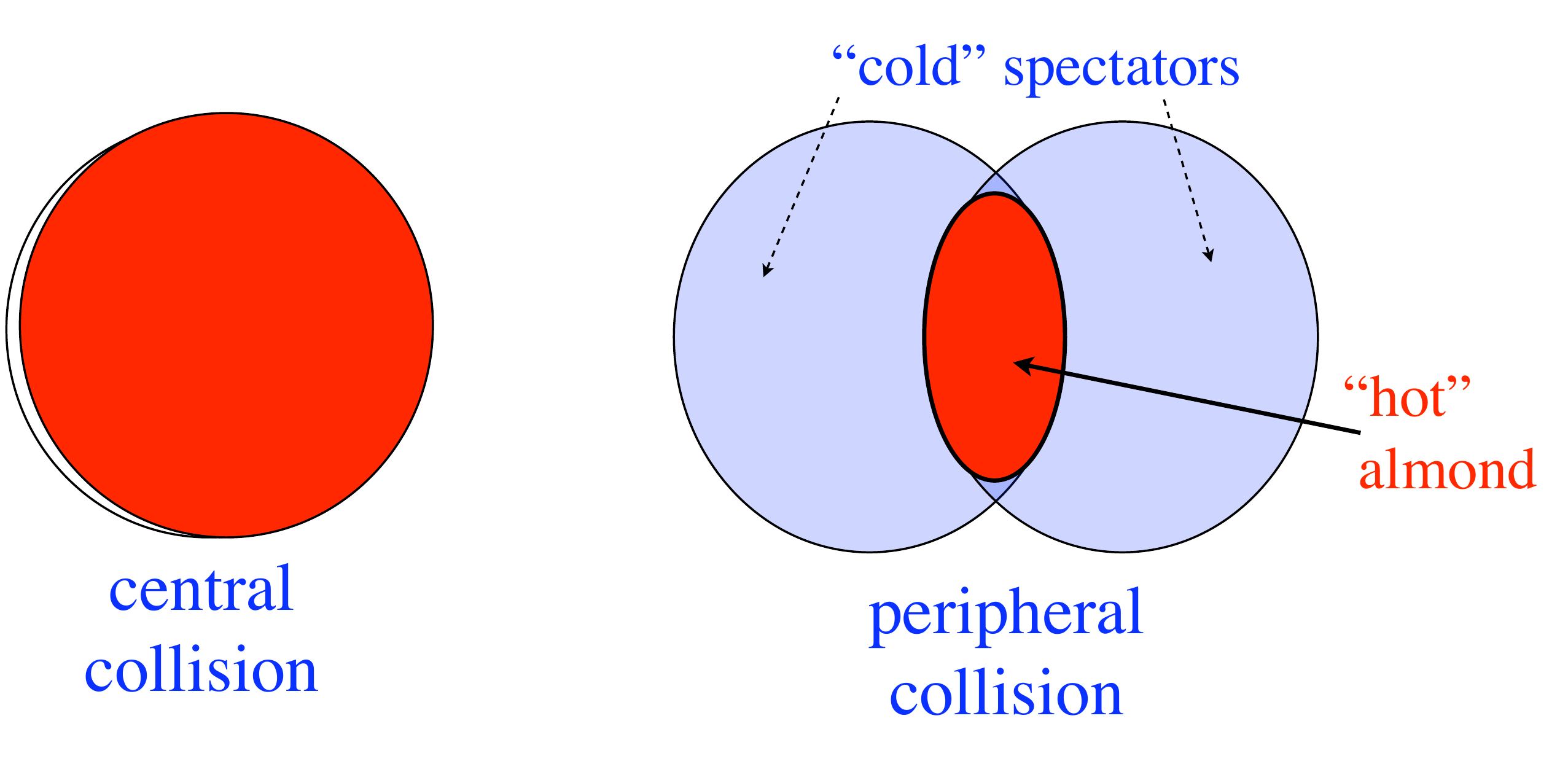}
\end{center}
\caption{Central versus peripheral collisions for large nuclei.}
\label{centralvsper}
\end{figure}

At RHIC, the particles are spread out over $\pm 5$ units of rapidity.
At zero rapidity there are $\sim 900$ 
particles per unit rapidity, versus $\sim 600$
at the highest energies at the SPS.
This sounds like a large number, but in fact, it is not.  
The total number of particles in a central $AA$
collision should scale like $A$: $A^{2/3}$ for the area of one nucleus,
times $A^{1/3}$ in length as it goes through the other nucleus.
Starting with 
the total number of particles in a $pp$ collision at these
energies, and multiplying by $A$,
one finds that, proportionally, there are {\it only} about $30\%$ more
particles produced in $AA$ collisions than a trivial extrapolation
from $pp$.
This is a strong constraint
on the physics, as it shows that at these energies,
there is a {\it small} amount of entropy
generated in $AA$ collisions, relative to $pp$.  

There are two large experiments at RHIC: PHENIX \cite{phenix} and STAR
\cite{star}, each with about 400 people; and two smaller ones, BRAHMS
\cite{brahms} and PHOBOS \cite{phobos}, each with about 50.
An amusing but specious observation is that the
total number of experimentalists working on the associated experiments
nearly equals the particle multiplicity (per unit rapidity).  This increases
slowly, only logarithmically, with energy; the number of theorists
grows much slower, perhaps as the log of a log...

At RHIC, STAR \cite{star} and BRAHMS \cite{brahms}
have shown there is a narrow plateau in rapidity,
in which the multiplicity, $dN/dy$, and the average transverse momentum,
$\langle p_t \rangle$, of identified particles are both constant over
$\pm 0.5$ units of rapidity.  

Given the large transverse size of large nuclei, 
as illustrated in fig. \ref{centralvsper}
one can distinguish between
``central'' collisions, where the nuclei overlap completely, and
``peripheral'' collisions, where they only partially overlap;
the direction of the beam is into the page.
Experimentalists speak of the number of particpants in a collision:
for a central collision with $A\sim 200$, this is $\sim 400$.
The number of participants can be determined
down to about $\sim 30$, especially by using Zero Degree Calorimeters to
measure what goes down the beam pipe.  

\section{Soft particles: elliptic flow and nearly ideal hydrodynamics}

Numerical simulations on the lattice indicate that at zero quark chemical
potential, there is a crossover to a new regime at $T_c \sim 150 - 200$~MeV
\cite{lattice}.
It is natural to think that the most obvious signals for a new state of
matter would be from soft particles, whose characteristic transverse
momentum is of order $T_c$.

\begin{figure}
\begin{center}
\includegraphics[width=0.5\textwidth]{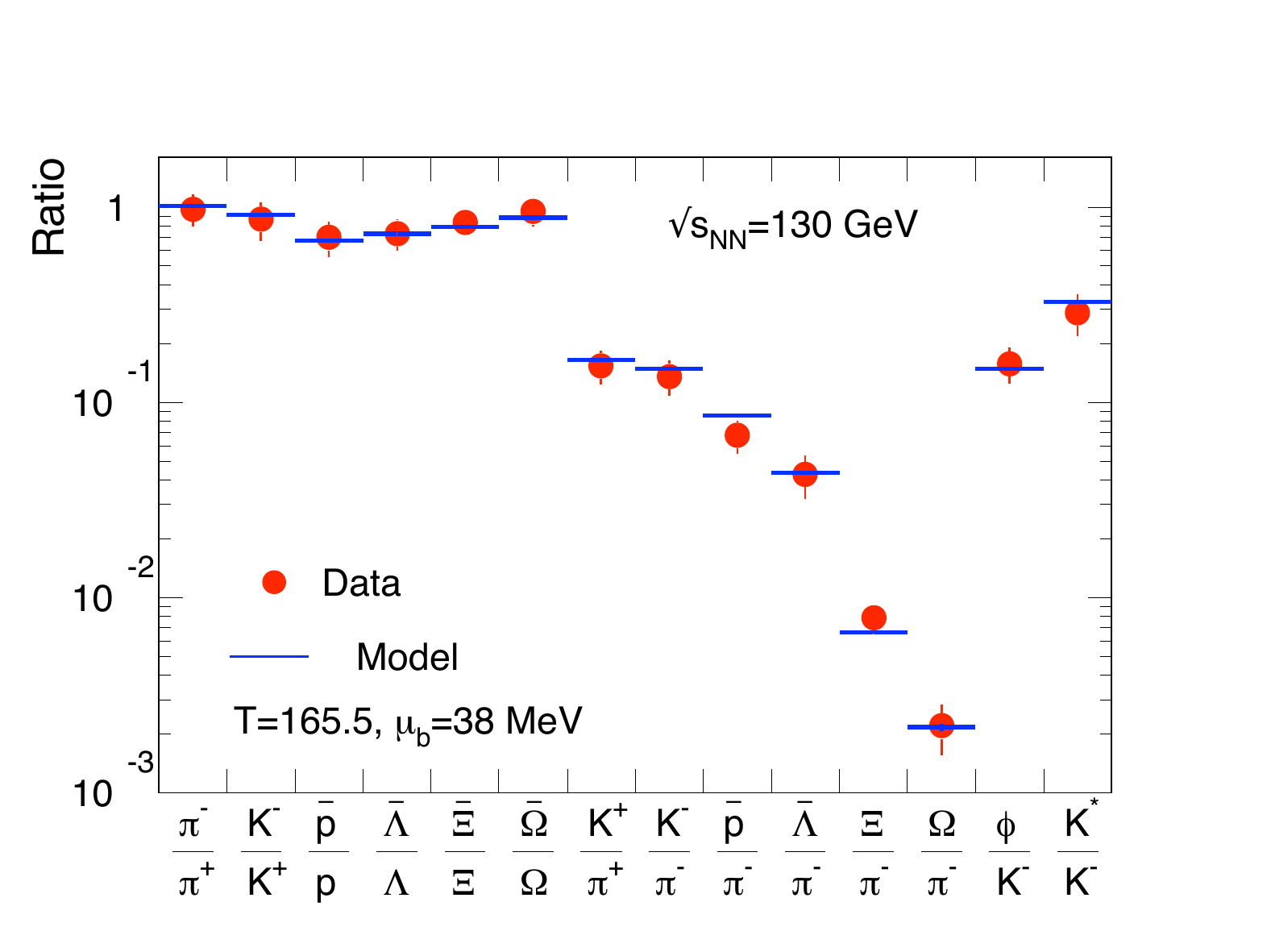}
\end{center}
\caption{Ratio of particle species, assuming chemical equilibrium.}
\label{chemeq}
\end{figure}

The first thing one can ask is about total particle multiplicities,
integrated over $p_t$.  This is illustrated in fig. \ref{chemeq} 
\cite{andronic}, which is a fit to over a dozen particle species with
only two parameters, a temperature $T_{\rm chemical} \sim 165$~MeV, and
a baryon chemical potential, $\mu_{\rm baryon} \sim 38$~MeV.  It does
not include short lived resonances, such as the $\Delta$, $\phi$,
$K^*$, {\it etc.}, but with a $\chi^2$ per d.o.f. of $4/11$, is 
an amazingly efficient summary of the data, using a trivial
calculation.  I remark that this is unlike analogous fits to
$pp$ or $e^+e^-$ collisions, where it is necessary to include 
other parameters which are not standard in textbook thermodynamics.  I 
stress that I do {\it not} claim that chemical equilibrium has been reached
in $AA$ collisions; theoretically, I do not know an unambiguous way of
verifying this.  Experimentally, though, overall ratios do appear to
look like it.

\begin{figure}
\centering
\includegraphics[width=.6\textwidth]{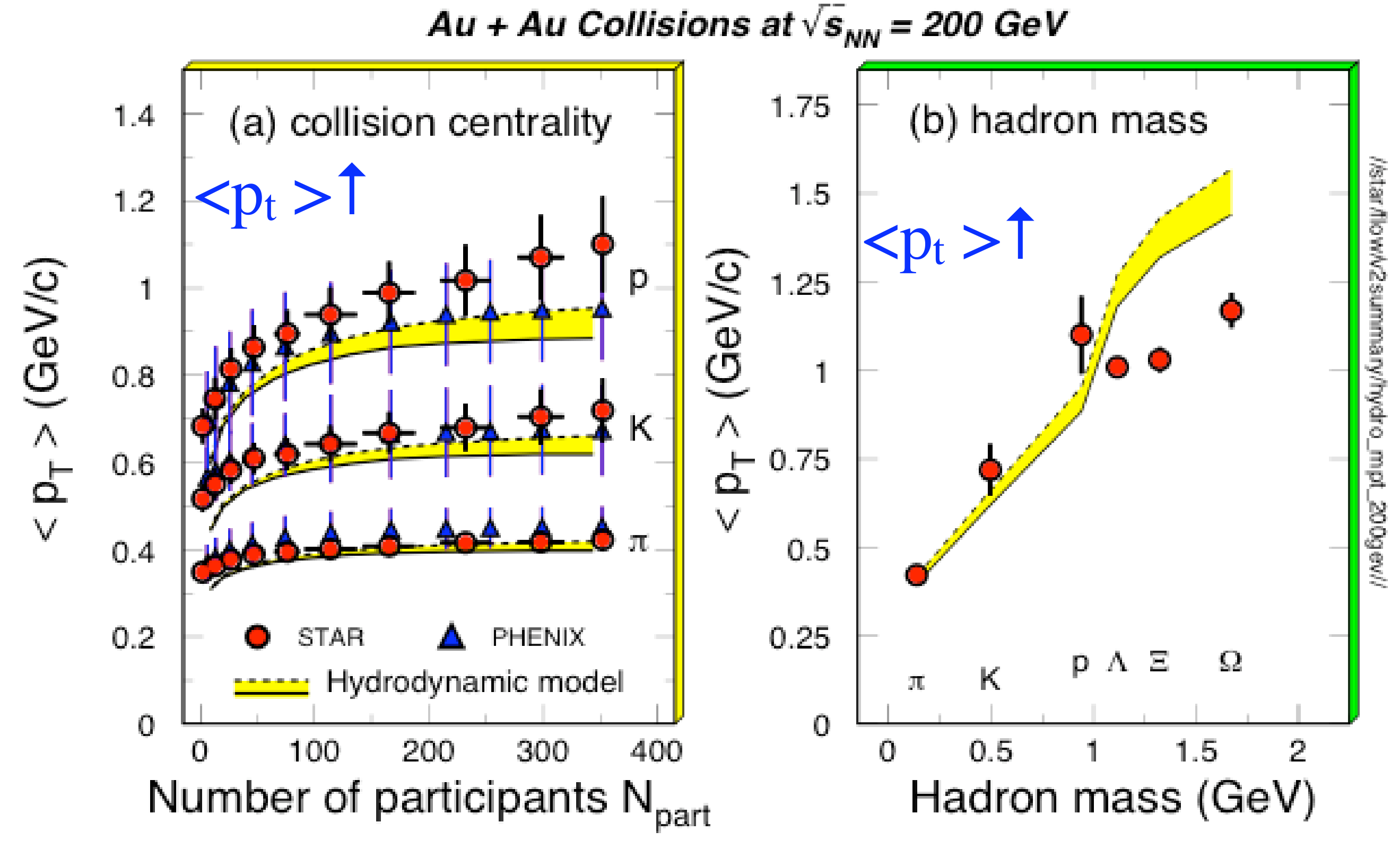}
\caption{Average transverse momentum, $p_t$, for different species.}
\label{avgpt}
\end{figure}

Instead of total multiplicity, integrated over $p_t$, the next thing one
can ask about is the average $p_t$, versus particle species.  This is
illustrated in the two figures of fig. \ref{avgpt}.

The figure on the left shows the change in the average $p_t$ for
pions, kaons, and protons, as one goes from $pp$ (on the left)
to the large nuclei in central $AA$, with $A \sim 200$ (on the right).
One sees a large increase in the average $p_t$ for kaons, and especially,
protons.  This is taken as evidence of radial flow in the collisions of
large nuclei: if a particle of mass $\rm m$ flows with a velocity
$\rm v$, its average transverse momentum should scale as 
$\langle p_t\rangle \sim  {\rm m v}$.  
Fits to the spectra indicate that one needs a flow velocity
${\rm v} \sim 0.6\, \rm c$.  The effect is more dramatic the
heavier the particle is, because light particles, such as pions, already
have an average velocity near the speed of light.

What I find striking about this figure, however, is that the average
momentum of pions does {\it not} increase significantly in going from
$pp$ to $AA$ collsions, with $A\sim 200$.  In a hydrodynamic description,
there is no reason for it to, but in the Color Glass model, the saturation
momentum $Q_s^2 \sim A^{1/3}$, and so one would expect the average $p_t$
to increase by a factor of $A^{1/6} \sim 2.4$.
One could easily
imagine having further increases in the average $p_t$ for kaons and protons
on top of that, due to radial flow.  But this doesn't happen: the average
$p_t$ for pions barely budges.  
Isentropic expansion decreases the average $p_t$ without increasing 
the multiplicity, so models of the Color Glass plus hydrodynamics fit 
this result, Fig. 5 of Ref. \cite{hirano}.  
It will be very interesting to see if this changes at the LHC.

One can then turn to the average $p_t$ of heavier species,
shown in the figure on the right hand side of fig. \ref{avgpt};
this plot is due
originally to Nu Xu.  It shows the average $p_t$ for central $AA$
collisions with $A \sim 200$ at the highest energies at RHIC.  
As seen in the figure on the left hand side, there is a
linear increase in the mean $p_t$ between pions, kaons, and protons; 
for heavier species, though, the $\Lambda$, $\Xi$, and $\Omega$, they
all appear to have nearly {\it constant} $p_t \sim 1.1$~GeV, like
that of the proton.  These
species are all baryons, but it is also found to be true of the $\phi$
meson.  

The usual explanation is that hadrons composed of strange quarks
decouple earlier.  Even if one assumes that strange
particles bunch up into colorless hadrons sooner, though,
I would expect that in a graph of $\langle p_t \rangle$ versus
mass, that there is one value of the slope of $\pi$, $K$, and $p$,
and another, smaller value, for strange particles.  Instead, it appears
as if starting with the proton, that it and all heavier hadrons
are emitted with essentially {\it constant}
$p_t$.  This is very difficult to understand from any hydrodynamic
description, where $\langle p_t \rangle \sim {\rm m v}$.

\begin{figure}
\centering
\includegraphics[width=0.4\textwidth]{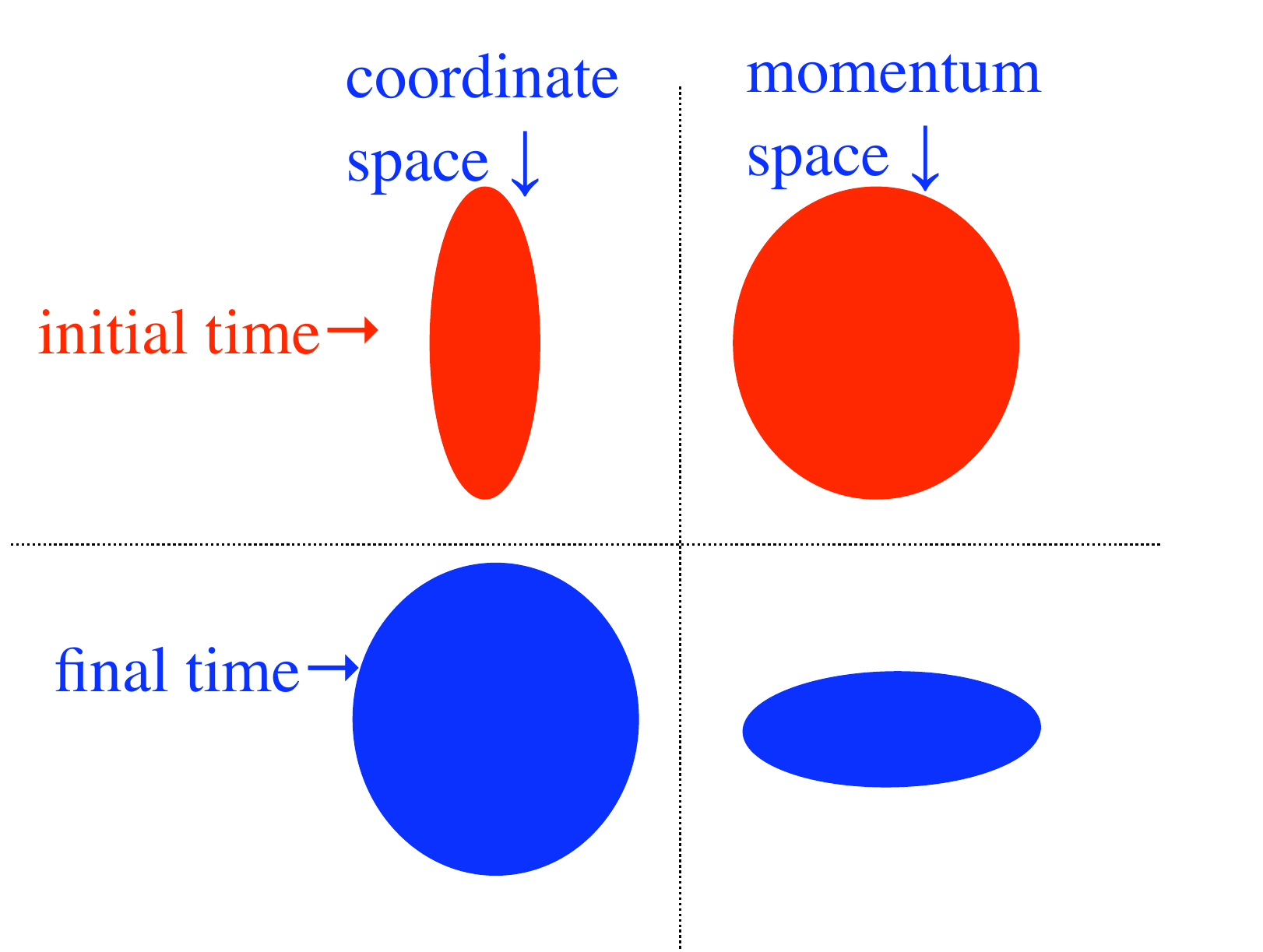}
\caption{Elliptic flow in a peripheral collision: evolution in both
coordinate and momentum space}
\label{elliptic}
\end{figure}

A fundamental quantity to measure in heavy ion collisions is that of 
elliptic flow.  To understand this, consider the hot ``almond'' of
the overlap region in a peripheral collision, fig. \ref{centralvsper}.
This is shown also in the upper left hand corner of fig. \ref{elliptic},
as a region in coordinate space.  The corresponding region in momentum
space is shown in the upper right hand corner of fig. \ref{elliptic}:
it is spherical, because by causality
particles can't start out knowing
the shape of such a large collision region.  
As the system evolves, and fields scatter off of one another,
in coordinate space the final distribution tends toward one which is spherical;
this is shown in the lower left hand corner of fig. \ref{elliptic}.  
At the same time, as the particles scatter, the 
distribution of particles in momentum space becomes distorted, into
an ellipse: particles along the $x$-axis, where the almond is narrow,
move a lot, while those along the $y$-axis, move less.  This is
characterized by the quantity
\begin{equation}
{\rm v}_2 
= \frac{\langle p_x^2 - p_y^2 \rangle}{\langle p_x^2 + p_y^2 \rangle}
\; .
\label{define_v2}
\end{equation}
This quantity is well defined and
so can be measured experimentally.  The main problem is determining the
reaction plane; {\it i.e.}, what are the $x$ and $y$ axes.  
One can also
define and measure higher moments, {\it etc.}

Nuclear physicists who work on collisions at lower energies are well
familiar with elliptic flow: then the two nuclei experience a lot of 
nuclear stopping, form a big blob that lasts a long time, and thus 
naturally transform the initial anisotropy in coordinate space
into one in momentum space.

At high energies, however, the mere existence of elliptic flow tells one
that there are significant interactions in $AA$ collisions.  The great
question about $AA$ collisions at high energies is whether there is
{\it anything} interesting beyond $A$ times a $pp$ collision.  Especially
in an asymptotically free theory, it is certainly conceivable that the 
particles, while originally in a almond, just free stream isotropically.
In this case, there would be {\it no} significant ${\rm v}_2$ generated.  

\begin{figure}
\centering
\includegraphics[width=0.7\textwidth]{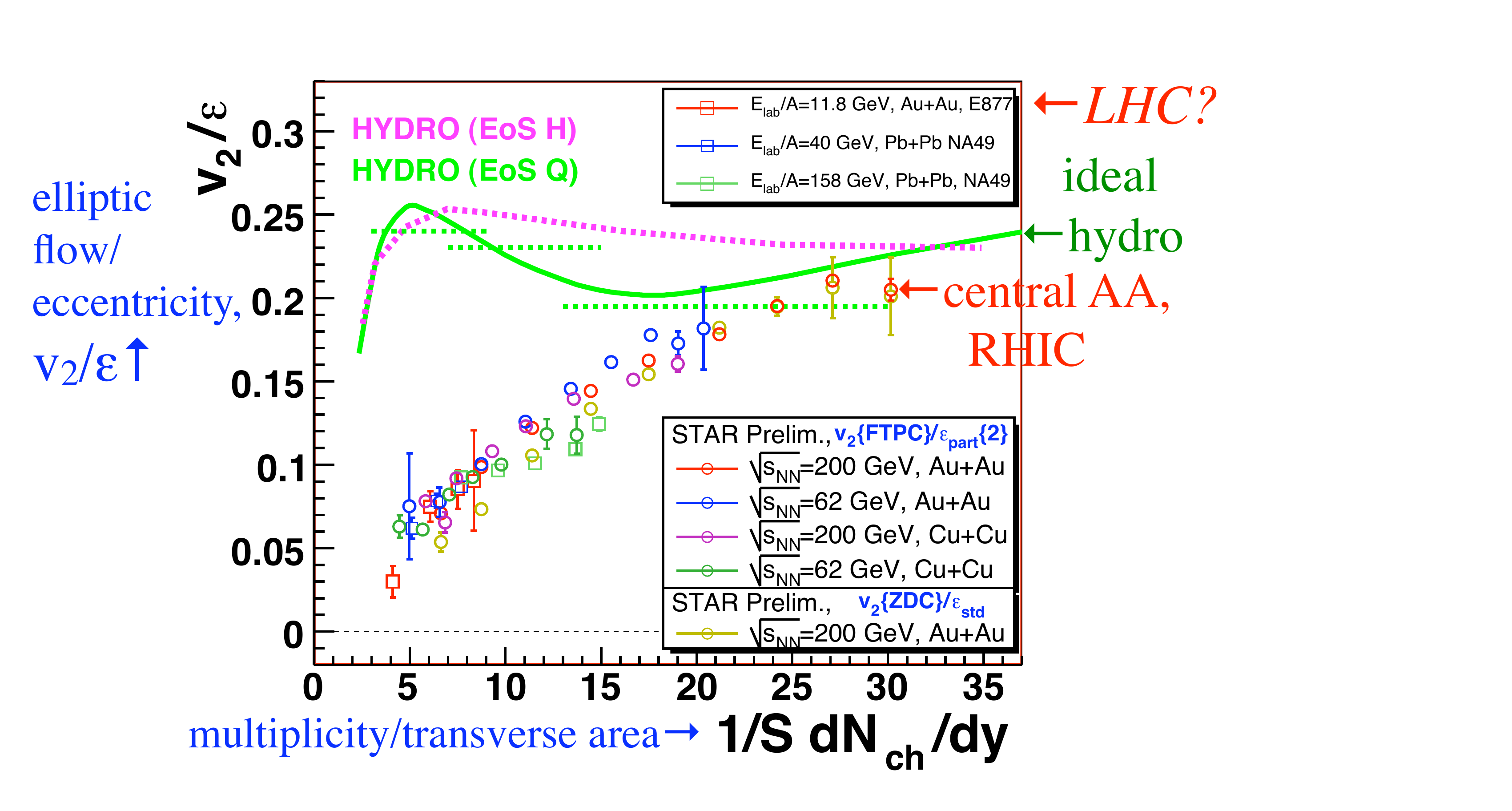}
\caption{Elliptic flow versus multiplicity}
\label{flowmult}
\end{figure}

One way of computing elliptic flow is to use a hydrodynamic description.
Given the large particle multiplicities, to zeroth order such a
description is reasonable, as hydrodynamics is a simple way of encoding the
conservation of energy and momentum in a causal manner.  Hydrodynamics
requires an equation of state; this one can take, for example, from
numerical simulations on the lattice \cite{lattice}.  
It is also necessary to specify
the transport coefficients of the medium, such as the shear and bulk
viscosity.  For a relativistic medium, there are other transport coefficients,
but we concentrate on the shear viscosity, 
as that appears to be largest and most important.  

Shear viscosity is familiar from the non-relativistic example of two
parallel plates, in the $x$ and $z$ planes,
separated by some distance in $y$.
If one plate is held fixed, and the other is moved with constant
velocity along the $x$-direction, then 
the shear stress is proportional to the viscosity
times the gradient of the velocity in $y$.  That is, the more viscous 
the fluid, the harder it is to move one plate parallel to the other.

The simplest thing one can do is to compute using {\it ideal} hydrodynamics,
assuming that the shear viscosity vanishes.  This is shown in
fig. \ref{flowmult}, which
shows the elliptic flow versus multiplicity in $AA$ collisions.
The elliptic flow is divided by the eccentricity, which allows
one to compare the collisions of copper nuclei, $Cu$, with $A \sim 60$,
to the largest nuclei, where $A \sim 200$.  
Plotting versus multiplicity (divided by
the transverse area) allows one to plots results
from energies at the AGS, SPS, and RHIC.
The basic point of this figure is that only for the collisions of
the largest nuclei, at the highest energies, that agreement between
data and (nearly) ideal hydrodynamics is found.
a nearly ideal hydrodynamics agrees with the data.
The best fit to the data is obtained with an
equation of state that includes a phase transition to a deconfined phase.

Hydrodynamics predicts both single particle distributions (versus
$p_t$) and elliptic flow.  It is found that 
elliptic flow provides a strong constraint
on the ratio of
the shear viscosity, $\eta$, to the entropy density, $s$ \cite{luzum}:
\begin{equation}
\eta/s \approx 0.1 \pm 0.1 ({\rm theory}) \pm 0.1 ({\rm experiment}) \; .
\label{boundeta}
\end{equation}
The experimental errors arise
from uncertainty as to the direction of the event plane; there are many
sources of error from theory.  The value quoted is for $\eta/s$, because
this enters naturally in hydrodynamics, and is related to an inverse mean
free path.

\begin{figure}
\centering
\includegraphics[width=0.4\textwidth]{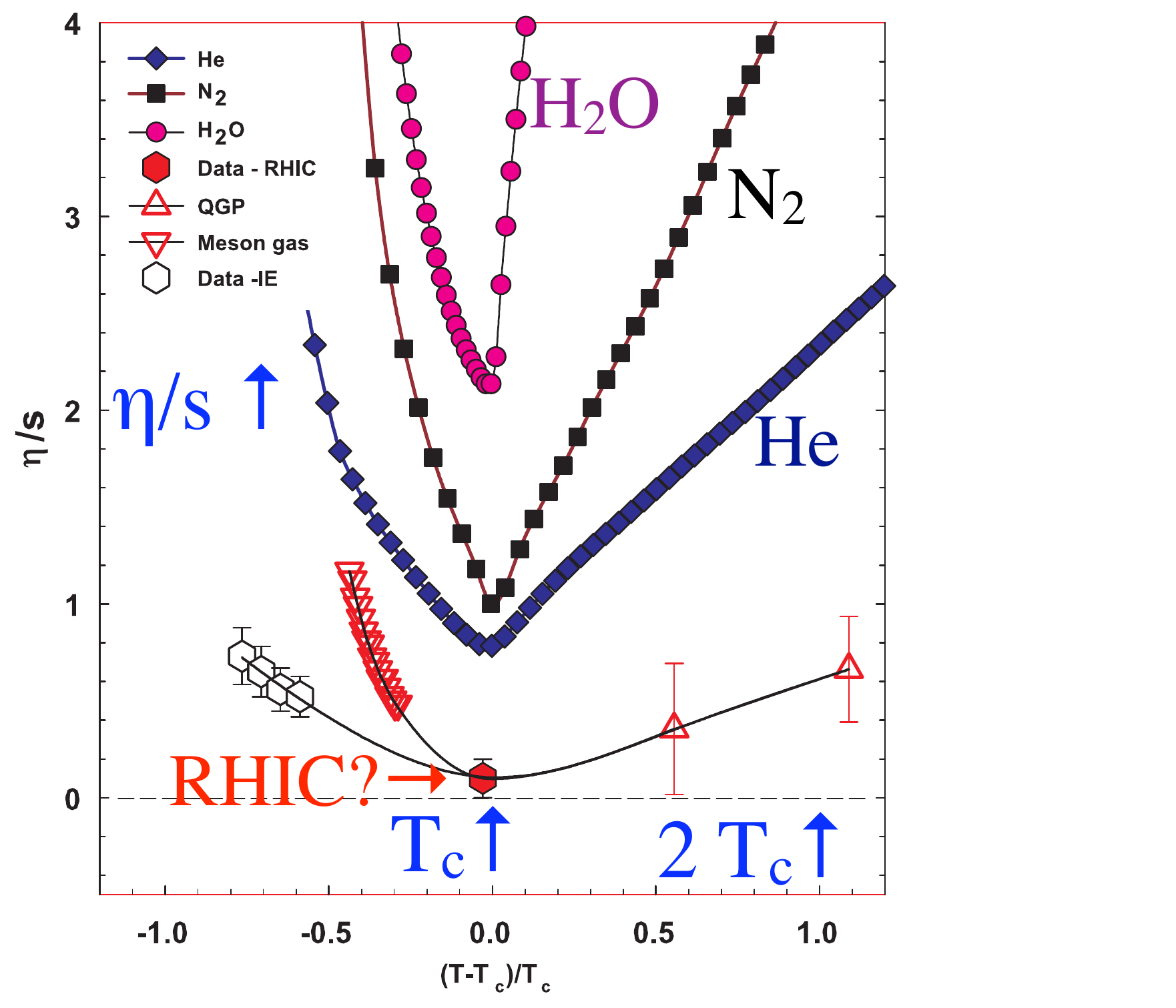}
\caption{Shear viscosity in various non-relativistic systems}
\label{shear}
\end{figure}

A comparison to various
non-relativistic systems is given in fig. \ref{shear}
\cite{lacey}.  The quantity
plotted is again $\eta/s$, but for non-relativistic systems, $s$ doesn't
change significantly near $T_c$, unlike for QCD, where it drops
dramatically \cite{lattice}.  Taking this ratio
does eliminate a trivial dependence on the overall number of the degrees of
freedom.

Even given the large error bars 
in eq. \ref{boundeta}, this is an
{\it extremely} small value for $\eta/s$.  (The points from a hadronic
gas and the QGP in fig. \ref{shear} are theoretical extrapolations.)
The value at RHIC is almost an order of magnitude smaller than
the smallest value for non-relativistic systems, which is liquid $He$.  
Thus RHIC produces ``the most perfect fluid on earth''.

As a transport coefficient, the shear viscosity vanishes in the limit
of weak coupling, as $\eta \sim T^3/\alpha_s^2$, where $T$ is the
temperature, and $\alpha_s$ the QCD coupling constant.  
The fact that $\eta$ is inversely proportional to a coupling constant
sounds peculiar, but it's not.
Transport coefficients measure how quickly a system,
perturbed from thermal equilibrium, goes back.  It takes longer for
a weakly coupled system, than a strongly coupled system, because the
particles interact less.
Technically, it is easiest computing $\eta$ from a Boltzmann equation.
There one finds that $\eta$ is the ratio of a source term (squared),
divided by a collision term: for small $\alpha_s$, the source term
is of order one, and the collision term $\sim \alpha_s^2$.
I do not discuss values of $\eta$ is weak coupling.  To date,
one cannot reliably compute either $\eta$ or the entropy near
$T_c$.  The situation is not hopeless, though 
\cite{reviews,hidaka,sonreview,brigante,susyeta}.

Since $\eta \sim 1/\alpha_s^2$, a small value for $\eta$ suggests
that the QCD coupling is very large near $T_c$.  This is part of the
motivation for what is known as a ``strong'' QGP \cite{reviews}.
One case where one can compute at infinite coupling is for a theory
with ${\cal N}=4$ supersymmetry and an infinite number of colors, 
where $\eta/s = 1/4 \pi$ \cite{sonreview}.
This is conjectured to be a universal bound, but string theory
provides examples which are $16/25$ smaller, and may
be the true bound \cite{brigante}.

As illustrated in fig. \ref{flowmult}, the
really interesting question is what the elliptic flow will be like
at the LHC.  Straightforward extrapolations of ideal hydrodynamics
can be done, and predict a large increase in ${\rm v}_2$
\cite{lhchydro}.  In this, there
appears to be real dichotomy.  
In a strong QGP \cite{reviews}, 
if the plasma is strongly coupled near $T_c$, at
RHIC, shouldn't it remain so at the higher temperatures at
the LHC?  Another example is provided by 
${\cal N}=4$ gauge theories:
by modifying the theory, they can be adjusted to fit
the pressure, as computed from numerical simulations on the lattice
for three colors \cite{lattice}, 
down to $T_c$ \cite{susyeta}.  In {\it all} of these models,
however, $\eta/s$ remains small, $= 1/4 \pi$.  

In contrast, as shown in fig. \ref{shear},
non-relativistic models universally show that while the
shear viscosity
has a minimum at the critical temperature, that it also increases
{\it away} from $T_c$.  The question is really, is the QGP
like $He$, where the increase from $T_c$ to $2 T_c$ is only a factor of
two, or like $H_2 0$, where it is an order of magnitude?  
A weak coupling analysis of a ``semi''-QGP suggests
that a large rise in $\eta/s$ is possible as $T$ increases from
$T_c$ \cite{hidaka}.

Measurements of the elliptic flow at the LHC will tell us from day one
of running $AA$ collisions.  I note that detailed theoretical predictions
in non-ideal hydrodynamics need to be carried out, since even if collisions
at the LHC start out in a highly viscous regime, at say $\sim 2 T_c$,
one still cools into a system which has a small viscosity near $T_c$.

\begin{figure}
\centering
\includegraphics[width=0.4\textwidth]{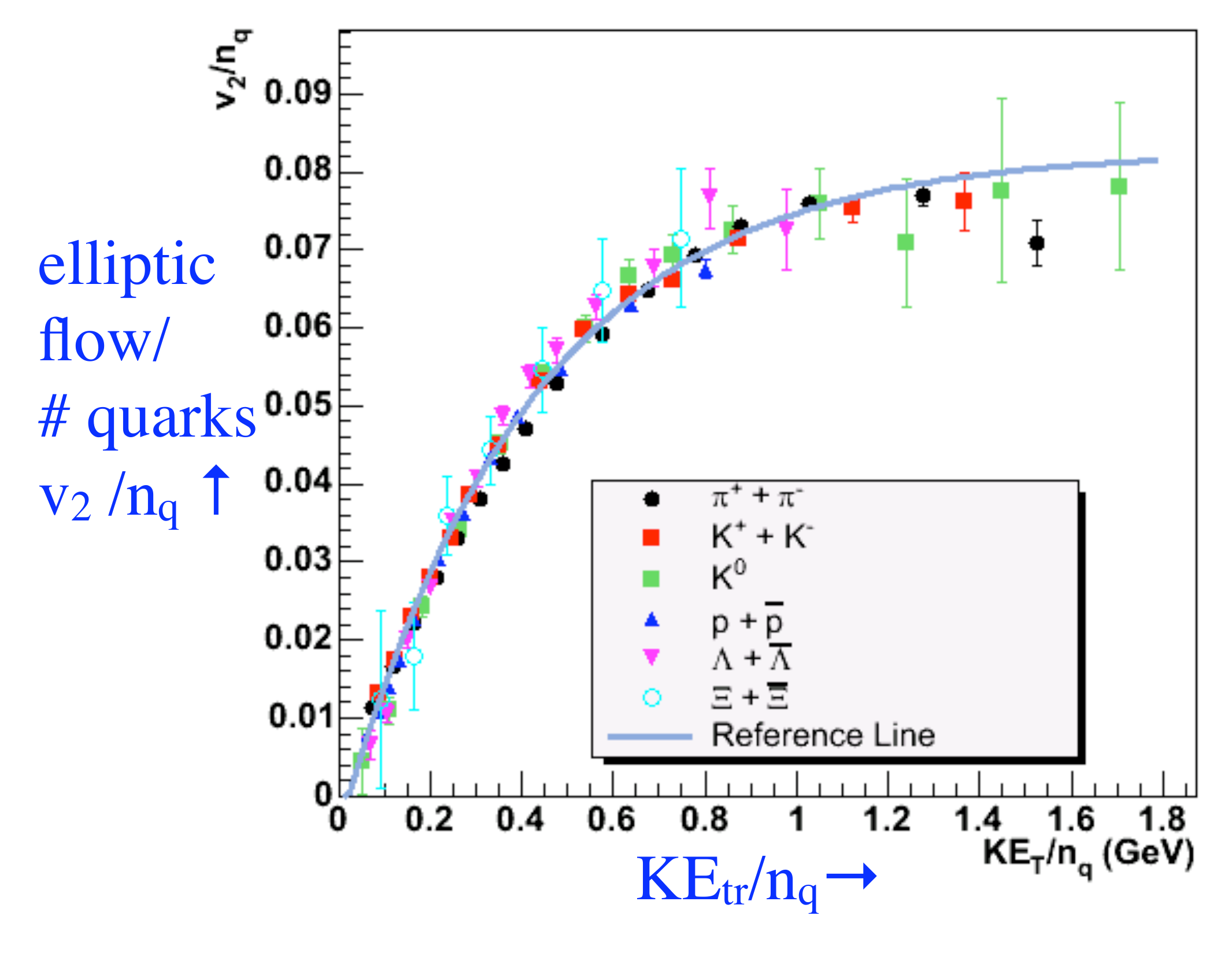}
\caption{Elliptic flow per quark, versus
the transverse energy per quark.}
\label{flow_energy}
\end{figure}

Returning to experiment, in fig. \ref{flow_energy}
I show a plot of the elliptic flow per quark, versus the transverse energy
of a hadron, per quark.  By per quark, I simply mean that one divides by
two for a meson, and three for a baryon.  This shows that at low
$p_t$, there appears to
be a universal scaling of elliptic flow for {\it all} particle species.
Dividing by the number of quarks in the hadron is reasonable,
but it is astounding that the correct variable to plot against is
the kinetic energy (and not, say, the transverse momentum; then one does
not find a universal curve).  This is typical of the results from RHIC:
there are many results which are simply totally unexpected, and hint at
some universal mechanism(s), which we do not yet understand.

\begin{figure}
\centering
\includegraphics[width=0.4\textwidth]{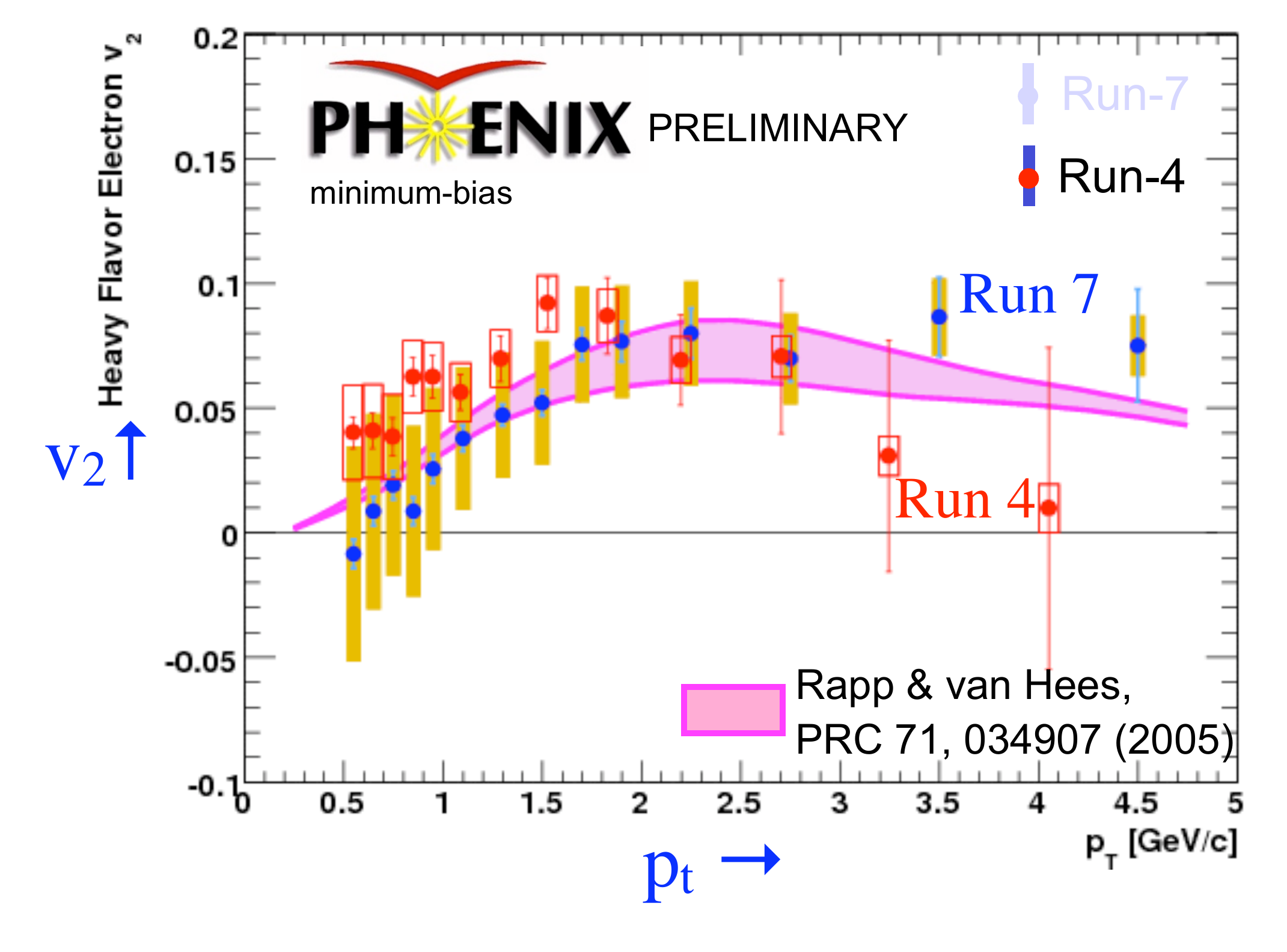}
\caption{Elliptic flow for charm quarks}
\label{flow_charm}
\end{figure}

One can also ask about the elliptic flow of heavy quarks.  Here experiment
uses single electrons, which arise from the decay of a charm quark, to
tag their flow.  Now theoretically, one would expect that heavy quarks
would {\it not} flow as easily as light quarks: it should take heavy
quarks longer to thermalize, and they should interact in a characteristically
different manner.  Instead, as shown in fig. \ref{flow_charm}, the
elliptic flow for charm quarks appears to be just as large as that of
light quarks!

This is one of the truly astounding results from RHIC.  As we shall see
again in the next section, heavy quarks appear to interact much more
strongly with the ``stuff'' in central $AA$ collisions than we would have
expected: $AA$ collisions are manifestly not a trivial
superposition of $pp$ collisions.  

\section{Hard particles: suppression and the ratio $R_{\rm AA}$}

\begin{figure}
\centering
\includegraphics[width=0.6\textwidth]{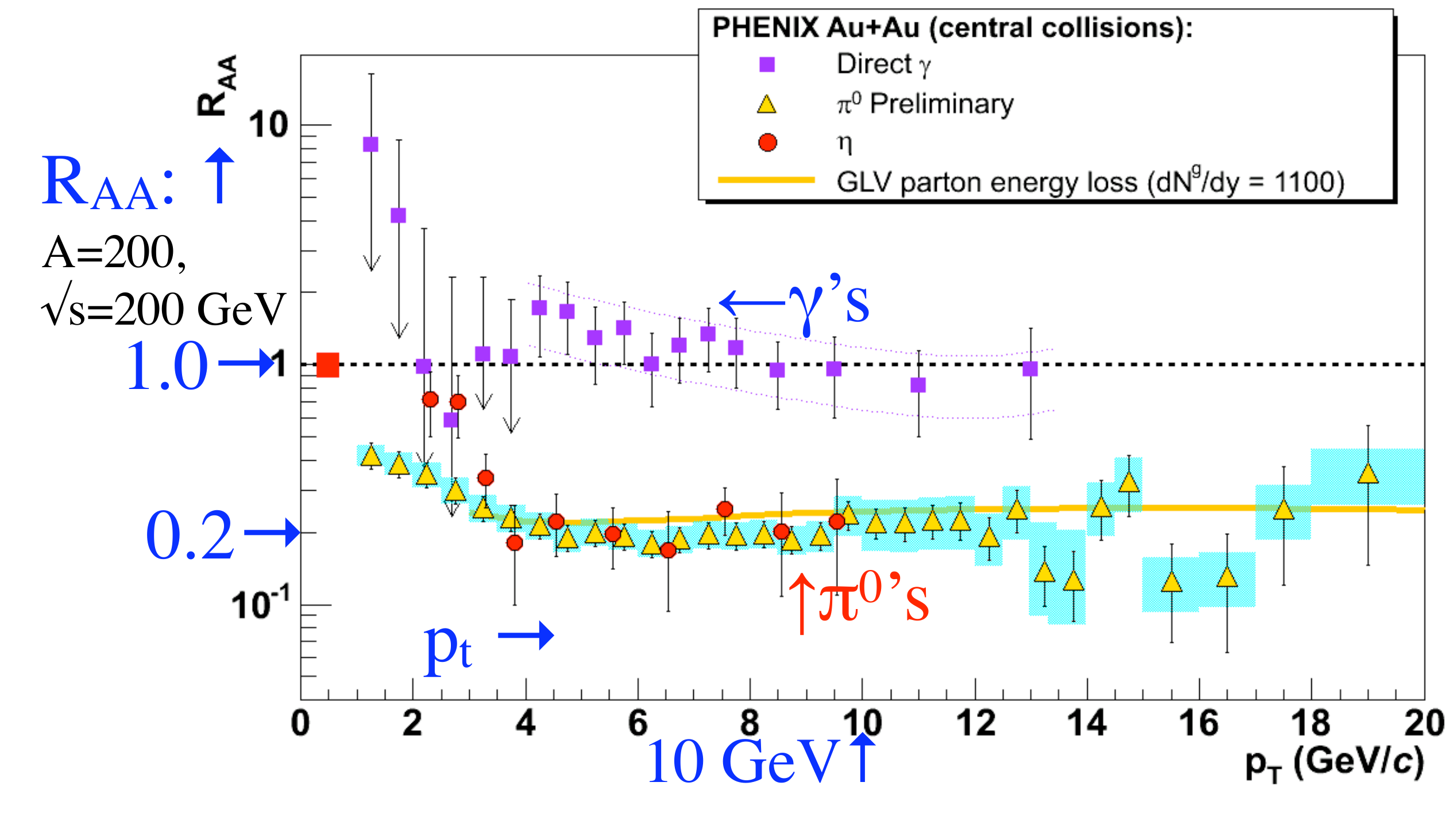}
\caption{The ratio $R_{\rm AA}$ for photons and pions.}
\label{rAA}
\end{figure}

One of the great experimental surprises of RHIC is that while most of
the particles are down at low $p_t$, the clearest signs for something
new in central $AA$ collisions comes from high momentum, $p_t > 2$~GeV.
This is typically referred to in the high energy nuclear community as
``jets'', but is far lower in energy than what most high energy
physicists are used to.  Consequently, I eschew this term, and just
refer to hard particles.

A basic quantity is the ratio $R_{\rm AA}$: this is the ratio of the
number of particles in a central $AA$ collision
to that in $pp$, both measured at the {\it same} $p_t$ (and rapidity):
\begin{equation}
{\rm R}_{\rm AA}(p_t) =
\frac{ \# \; {\rm particles \; in \; central \; AA }(p_t) }
{ {\rm A}^{4/3}\; \# \; {\rm particles \; in \; pp} (p_t)}.
\label{rAAeq}
\end{equation}
The crucial question is how one normalizes.  As I discussed above,
soft particles scale as $A$.  For hard collisions, the number of binary
collisions is $A$, from the incident nucleus,
times $A^{1/3}$ from the
width of the target, or $A^{4/3}$.  This is only approximate;
experimentally, this is modeled by Glauber and Monte Carlo calculations.

However, one doesn't need to understand (or believe) this normalization factor,
since one can directly appeal to experiment.  The ratio $R_{\rm AA}$
can be measured for any particle species.  In fig. \ref{rAA}, I show
the plot for photons and neutral pions.
Since photons only interact electromagnetically,
if the normalization is performed correctly, then $R_{\rm AA}$ should be
one.  While the error bars are large, $\sim 10\%$, this is true for
photons with $p_t > 2$~GeV.

In contrast,one finds that above $p_t \sim 2$~GeV, 
there are only about $20\%$ of the number of neutral pions expected.  
(Experimentally, at high $p_t$
it is easiest to pick out neutral pions, by looking for
two hard photons with the right invariant mass.)  This $20\%$ is
a {\it very} small number.  From fig. \ref{centralvsper}, even in a central
collision, there is a contribution from the surface; at least
half the hard particles emitted from the surface should escape without
interaction.  This is another reason why people speak of a 
strong QGP at RHIC \cite{reviews}.

Indeed, the really surprising thing is that $R_{\rm AA}$ is so {\it flat}
to such a high $p_t$.  It is easy to imagine that effects in a medium
would suppress hard particles: they will scatter off of the medium, lose
energy, and so emit more soft particles.  Theoretically, this
is known as energy loss \cite{reviews}.  But at high enough
$p_t$, scattering off of the medium should go away.  Fig. \ref{rAA}
shows that this isn't true for neutral pions with a $p_t$ as high as
$20$~GeV!  Eventually, $R_{\rm AA}$ must go back up
to one, or one will question whether it is correctly normalized.

It is reasonable to ask if this suppression is due to some
initial state effect in nuclei.  Here measurements
in $dA$ collisions were crucial: the normalization changes to $2A$, and
experimentally one observes not suppression, but enhancement
\cite{phenix,star,brahms,phobos}, with 
${\rm R}_{\rm dA} \approx 1.4 \pm 0.1$ at
$p_t \sim 3$~GeV.  This is due to what is known as the Cronin effect;
all that matters for us is that $R_{\rm dA}$ goes in the opposite direction
from $R_{\rm AA}$, and so $R_{\rm AA}$ is manifestly a final state effect.

\begin{figure}
\centering
\includegraphics[width=0.7\textwidth]{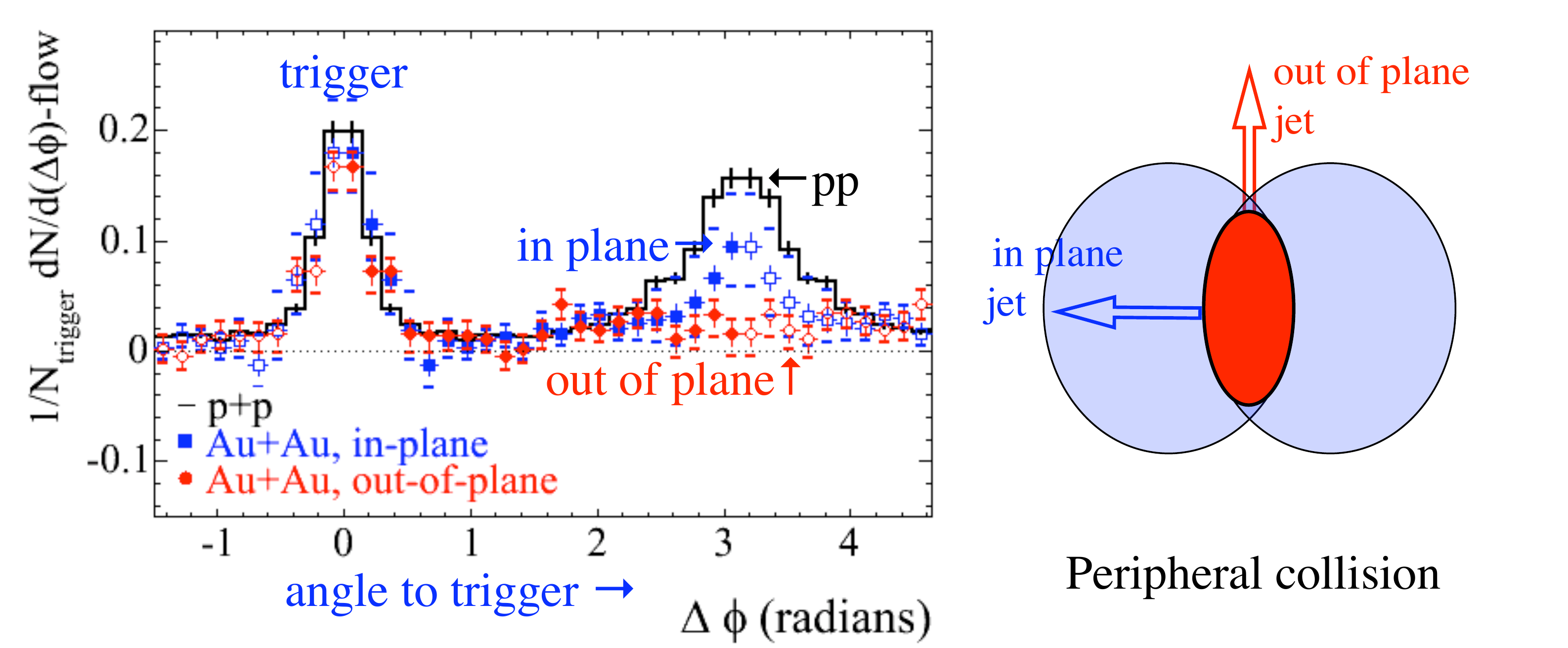}
\caption{Geometrical suppression of hard particles}
\label{geometry}
\end{figure}

The suppression of hard particles can also be observed on a purely
geometrical basis, as shown in fig. \ref{geometry}.  Consider
a peripheral collision, and trigger on a hard particle, with
$p_t : 4 \rightarrow 6$~GeV.  Then look for a hard particle on the away
side, $p_t > 2$~GeV, as a function of the angle to the trigger particle.
In $pp$ or $dA$ collisions, this is peaked at $180^o$.  Now in a
peripheral collision, one can look at a hard particle either in the plane of
the collision, or out of plane.  If the hard particle is in the 
reaction plane,
it goes a small distance through the ``hot'' almond, and a long ways
through the cold nuclear spectators.  If out of the plane, it goes a long
way through the almond, and little through the spectators.  Fig.
\ref{geometry} shows that when the hard particle is in the reaction plane,
one does see the away side particle at $180^o$;
when the hard particle is perpendicular to the reaction plane,
one doesn't see the away side particle.
That is, the more particles go through the almond, the more
the ``stuff'' there affects their propagation.
This is consistent with the small value of $R_{\rm AA}$.

\begin{figure}
\centering
\includegraphics[width=0.5\textwidth]{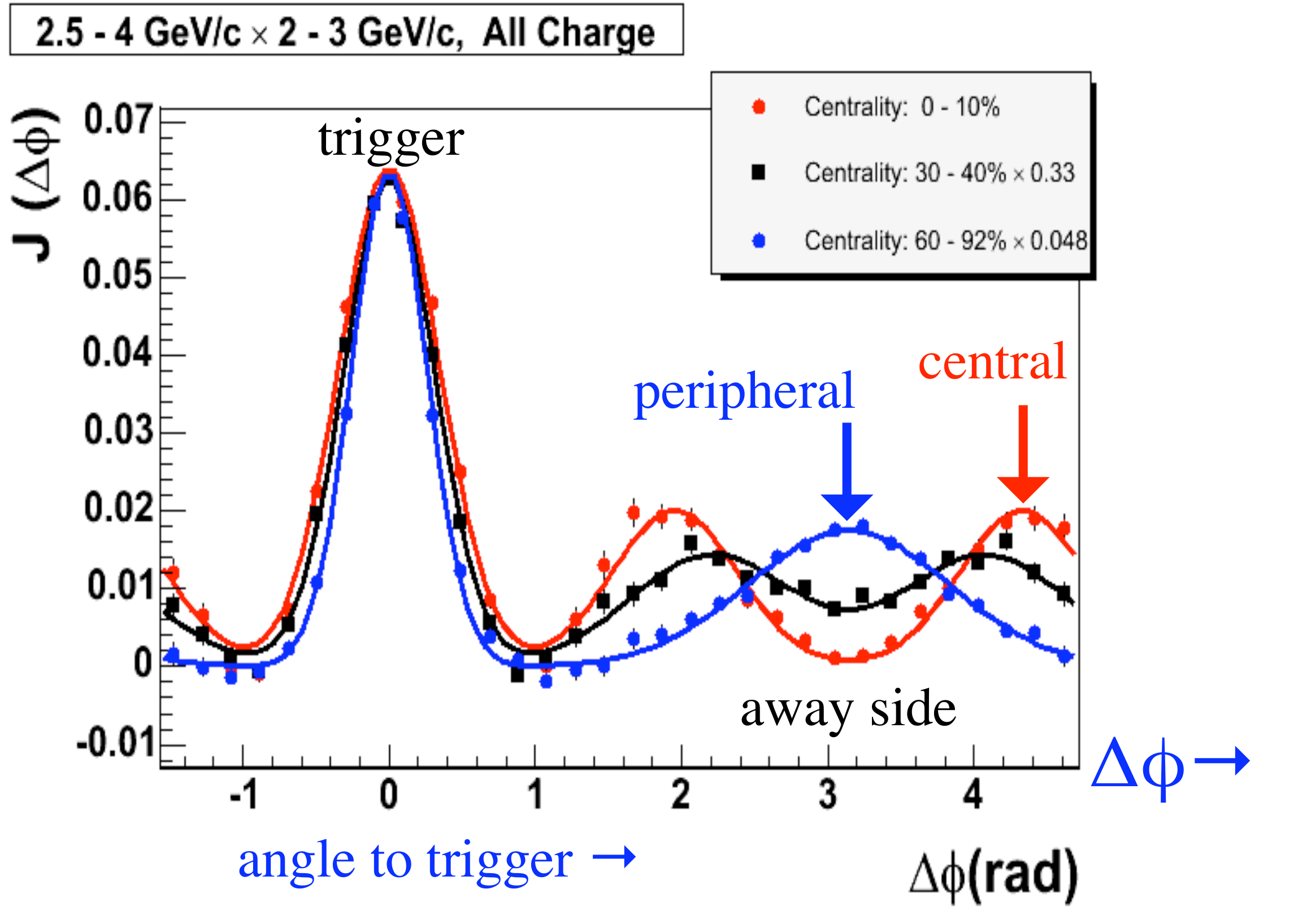}
\caption{Away side correlations for peripheral to central $AA$ collisions.}
\label{machcone}
\end{figure}

There is interesting structure seen in the angular correlations of
the away side particle.  Fig. \ref{machcone} shows results for a
trigger particle of $p_t : 2.5 \rightarrow 4$~GeV, and an away side
particle of $p_t: 2 \rightarrow 3$~GeV, integrated over all
angles to the reaction plane.  There are three curves shown, going
from most peripheral to most central.  How one defines centrality is
in this case secondary.  What one can see is that for peripheral collisions,
the angular distribution for the away side particle is peaked at 
$180^o$, as in a $pp$ collision.  For the most central collisions,
the angular distribution at $180^o$ is suppressed, as seen 
in fig. \ref{geometry}.
What one also sees, however, is 
an {\it enhancement} in the distribution of away side particle away
from $180^o$.  This looks very like Cerenkov radiation, or perhaps
a Mach cone in a medium \cite{reviews}.  
This is really a correlation between three particles, as has been
verified by both 
the PHENIX \cite{phenix} and STAR \cite{star} collaborations.

Especially with planned upgrades to RHIC, one will also be able to
measure correlations between a hard photon and a hard particle.
Measuring a hard photon
will tell one unambiguously what the incident energy of the hard
particle is, and so one will be able to understand the details of
how fast particles are affected by the medium in central $AA$ collisions.

\begin{figure}
\centering
\includegraphics[width=0.4\textwidth]{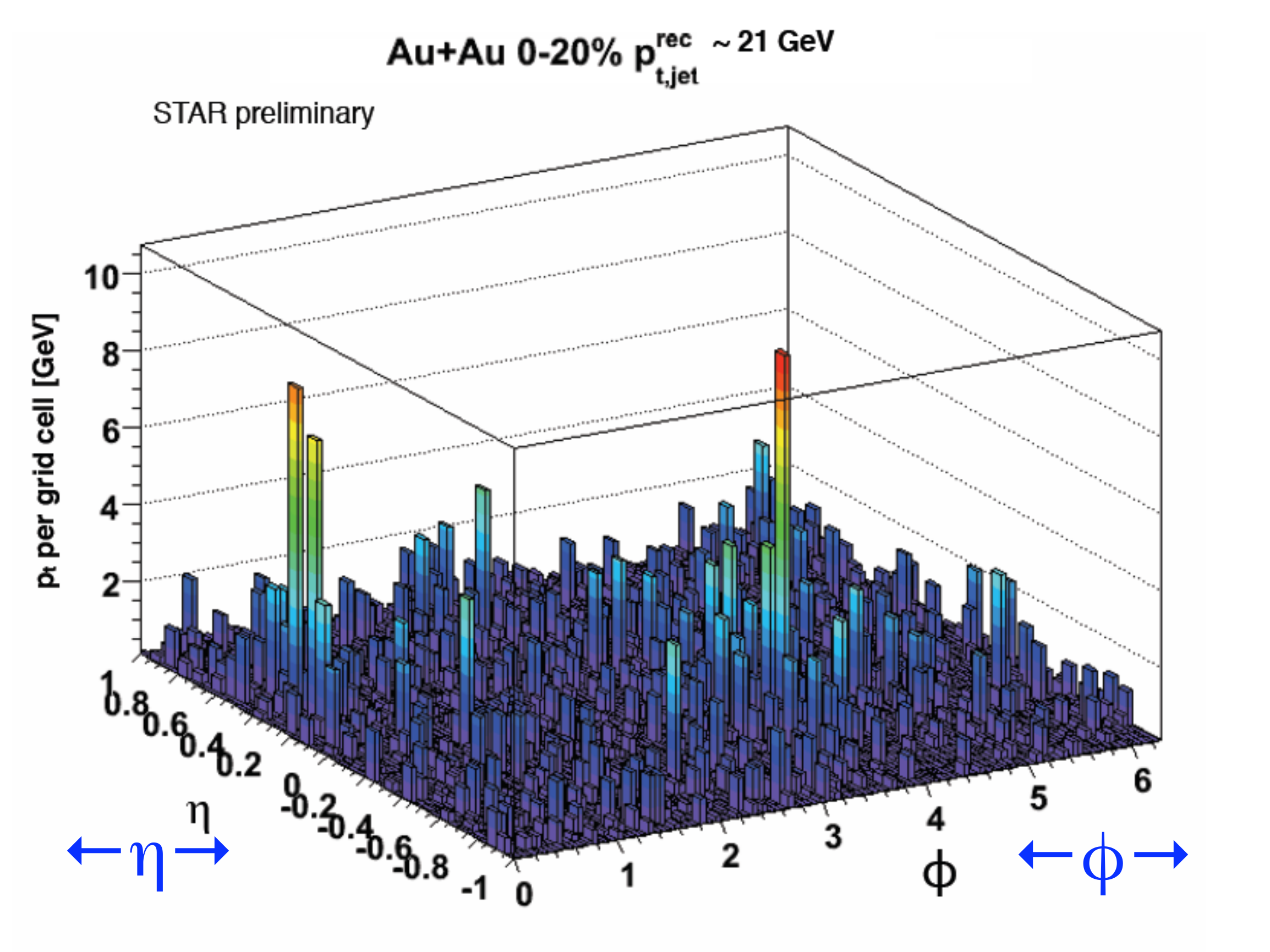}
\caption{A Lego plot of two jets in central $AA$, $p_t > 20$~GeV.}
\label{lego}
\end{figure}

All of these figures have triggered on ``hard'' particles with
relatively low $p_t$.
In fig. \ref{lego} I show a plot from
the STAR collaboration, which is a Lego plot familiar
in high energy physics.  The trigger is $p_t > 20$~GeV, for the most
central $AA$ collisions.  Even given the high multiplicity of particles
at low $p_t$, if the trigger is sufficiently high,
then jets just stick out.  
At LHC energies, true jets,
with transverse momenta of order $50$, $100$~GeV and higher,
will be (relatively) plentiful.  This will enable one to really pin down
the mechanism which is responsible for $R_{\rm AA}$ and the like.  

\begin{figure}
\centering
\includegraphics[width=0.5\textwidth]{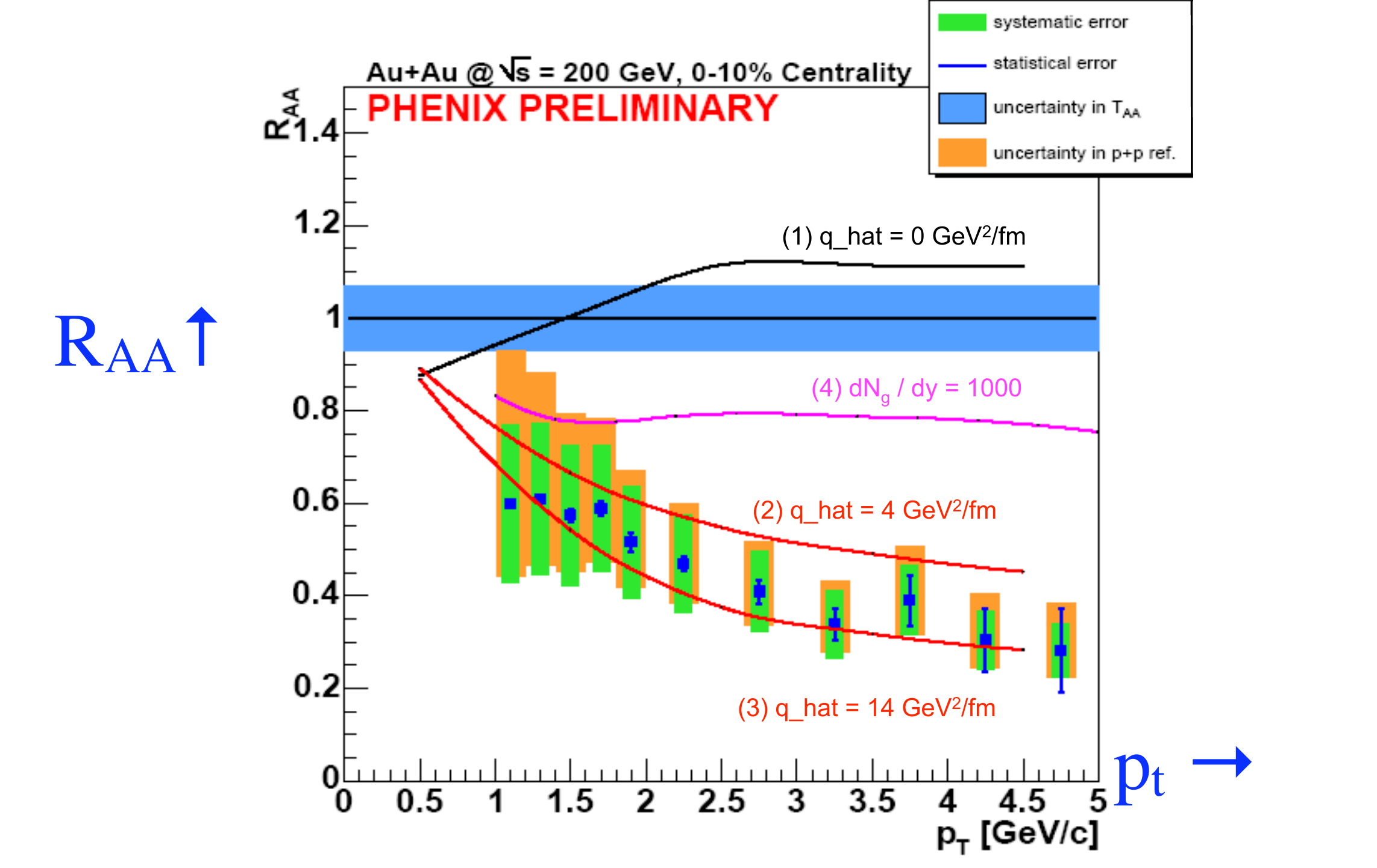}
\caption{The ratio $R_{\rm AA}$ for charm quarks.}
\label{charmRAA}
\end{figure}

One can form the ratio $R_{\rm AA}$ for any particle species.
In fig. \ref{charmRAA} I show the result for charm quarks from
the PHENIX collaboration.  Here one observes charm by measuring
direct electrons.  The mass of the charm quark is $\sim 1.5$~GeV,
and the temperature is something like $T_c \sim 200$~MeV \cite{lattice}.  In
perturbation theory, the scattering of a heavy quark is very different
from that of a light quark: emission of gluon radiation is suppressed
in the forward direction (``dead cone'' effect).  Even without detailed
calculation, it would be astonishing if one found that the behavior of
a heavy quark were anything like that of a light quark; one expects
that heavy quarks are not suppressed as much as light quarks, with
so $R_{\rm AA}$ is larger.

This is {\it not} what experiment shows: fig. \ref{charmRAA} shows that
for transverse momenta a couple of times the charm quark mass,
$p_t \sim 4$~GeV, that $R_{\rm AA} \sim 0.2$, like $\pi^0$'s!  
This is a remarkable result, and completely unlike any perturbative
understanding. Perhaps energy loss is not the whole story.

\section{Electromagnetic signals: dileptons and direct photons}

\begin{figure}
\centering
\includegraphics[width=0.5\textwidth]{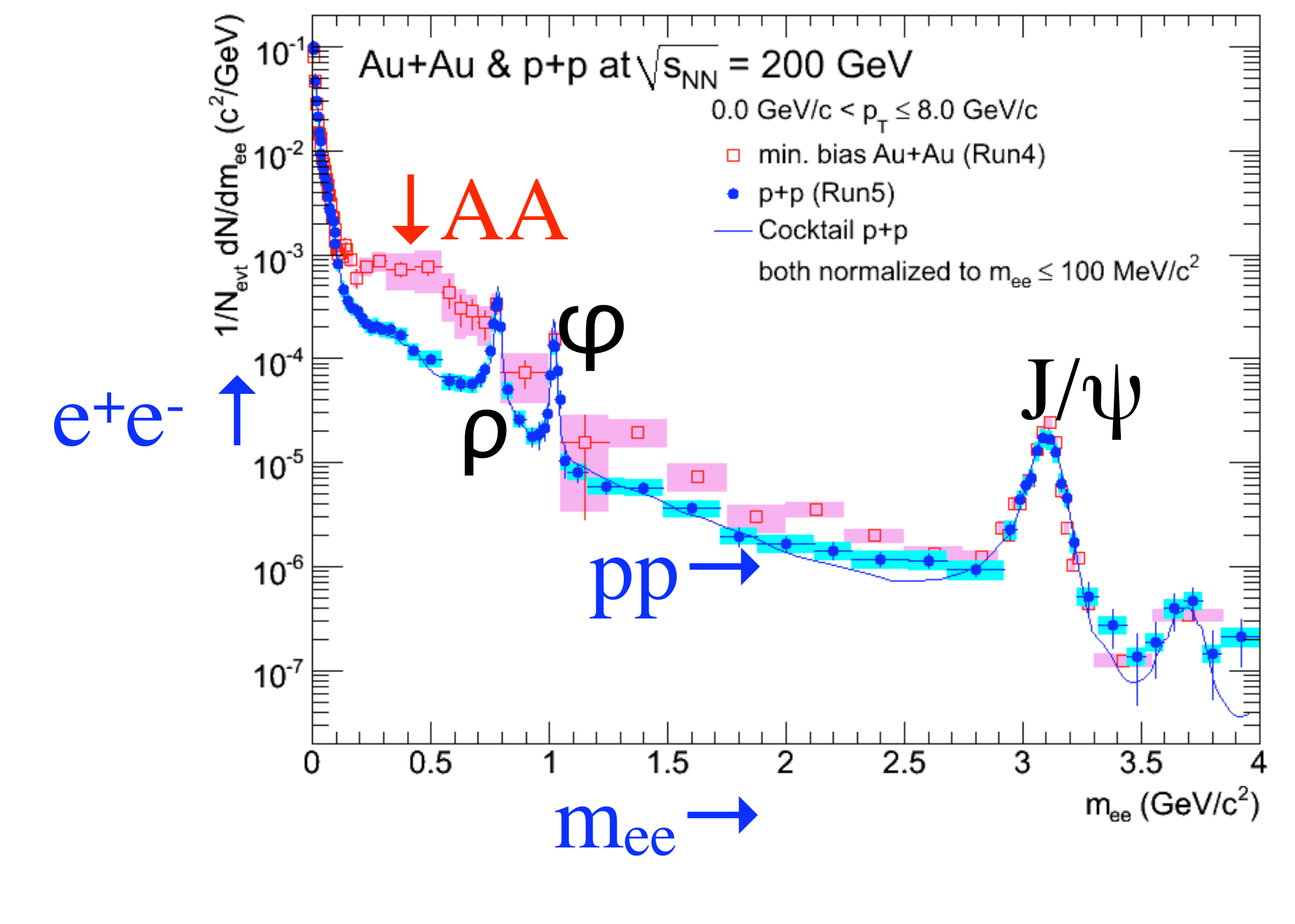}
\caption{Dilepton spectra for central $AA$ and $pp$ collisions.}
\label{dilepton}
\end{figure}

Since dileptons only interact weakly with a hadronic medium, they provide
essential insight into $AA$ collisions.  In fig. \ref{dilepton} I show
the dielectron spectra below the $J/\Psi$, as a function
of the invariant mass of the dielectron pair, $m_{\rm ee}$.  It is necessary to
normalize the spectrum from central $AA$ collisions to that of a
``cocktail'' from $pp$ collisions.

As seen in collisions at SPS energies, at RHIC energies there is a
striking excess in dileptons below the $\rho$ meson.  There is a smaller,
but still significant excess, above the $\rho$ meson as well.  Any
excess appears to have disappeared for dileptons above the $J/\Psi$.

A crucial question is whether the normalization to $pp$ collisions
is done correctly.  One can show that for 
the dilepton excess below the $\rho$ meson, for 
$150< m_{ee} < 750$~MeV, that the
excess first appears when the number of participants
is greater than $\sim 200$, and that it increases as the number of participants
increases.  This is dramatic evidence that the ``stuff'' created in
central $AA$ collisions is uniquely responsible for the excess at low
invariant mass.

\begin{figure}
\centering
\includegraphics[width=0.5\textwidth]{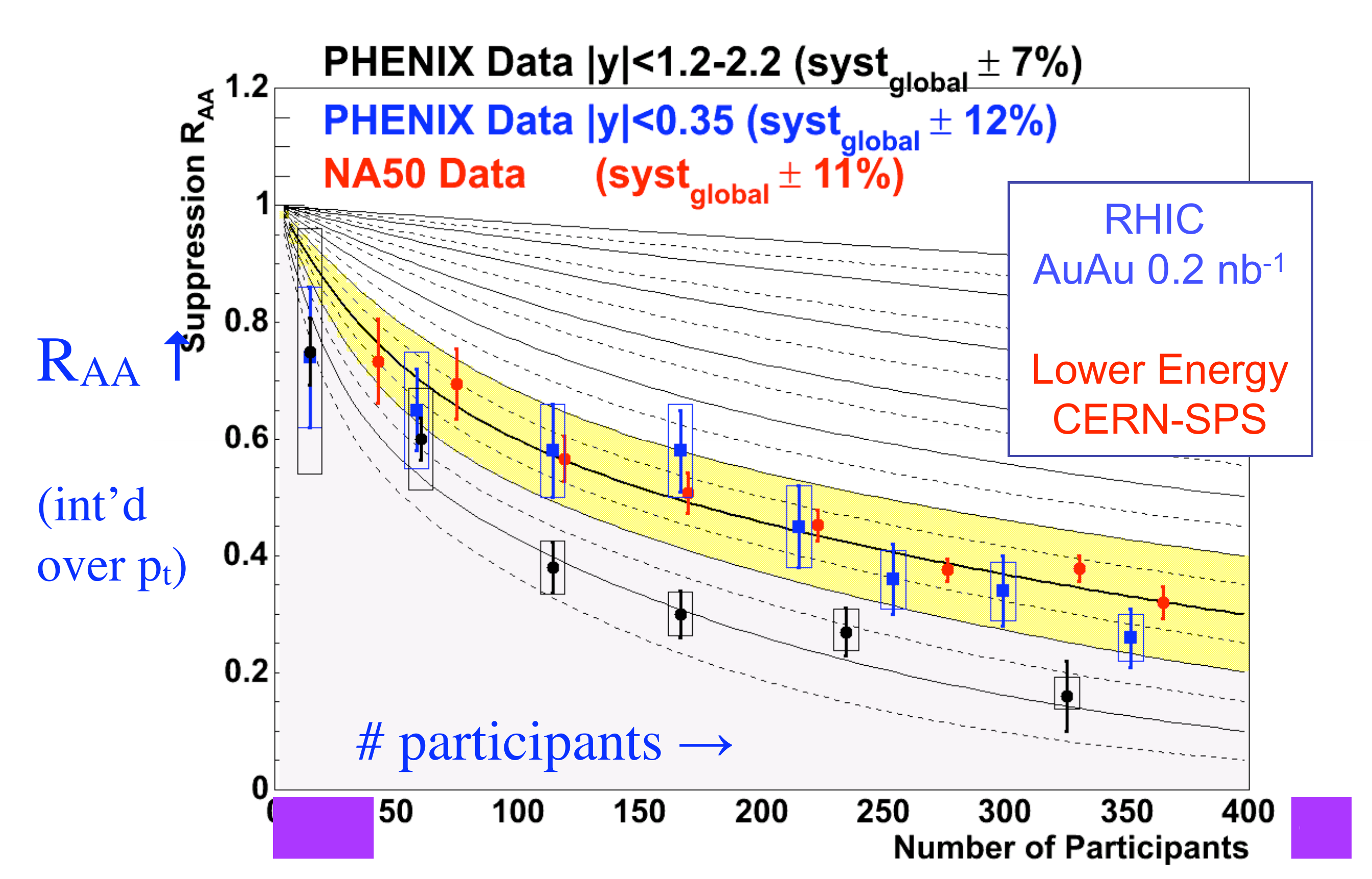}
\caption{The ratio $R_{\rm AA}$ for $J/\Psi$'s.}
\label{Jpsi}
\end{figure}

In fig. \ref{Jpsi} I show the ratio $R_{\rm AA}$ for $J/\Psi$ production
in central $AA$ collisions at both RHIC and the SPS.  When plotted in
this way, one finds that
the behavior at these two energies is essentially {\it identical}.  This
was absolutely unexpected.  Various theoretical models had predicted that
$J/\Psi$ production might be less at RHIC than the SPS, due to greater
scattering in a thermal medium, or greater, due
to regeneration.  But {\it no} model predicted exactly
the same behavior for $R_{\rm AA}$.

This year, PHENIX has also shown how low mass dielectron pairs can be
used to get direct photons from internal conversion \cite{photon}.
They see a clear excess for photon $p_t: 1 \rightarrow 3$~GeV, which they
fit to an exponential.  This gives a temperature for photon production
of
$T_{\rm photon} \sim 223$~MeV,
with statistical errors of  $\pm 23$~MeV and systematic errors of
$\pm 18$~MeV.
This is a fundamental result, and gives us a lower bound on the temperatures
at which the photons were produced.

\section{Summary}

The results at RHIC have conclusively demonstrated that central $AA$
collisions have produced matter at high energy density which is {\it very}
unlike that produced in $pp$ collisions at the same energy.  

There are numerous interesting phenomenon which I didn't have space to
cover: the baryon/meson enhancement at intermediate 
$p_t: 2 \rightarrow 6$~GeV; Hanbury-Brown-Twiss interferometry, which
shows ``explosive'' behavior; and the ridge in rapidity.  
I have emphasized that 
one of the most mystifying aspects of the data is that the behavior of
charm quarks --- as seen in their elliptic flow, 
and the ratio $R_{\rm AA}$ --- is essentially identical to that of light
quarks.  This is very difficult to understand theoretically.  This,
and the nearly ideal behavior of hydrodynamics, has given rise to the
suggestion that the region near $T_c$ is behaving unexpectedly:
either a strong \cite{reviews,sonreview,brigante,susyeta,lhchydro}, 
or maybe a semi- \cite{hidaka}, QGP.

Of course we eagerly await results for $AA$ collisions at the LHC.  
Collisions at the LHC will produce many more jets, and produce a medium
in which the temperatures are significantly (twice?) as high as at RHIC.
One might hope that LHC probes a perturbative (or complete \cite{hidaka})
QGP.  We will know very soon if LHC produces a nearly ideal fluid,
as at RHIC \cite{reviews,susyeta}, or one which is viscous \cite{hidaka}.
The study of bottom quarks will also be very interesting,
given the unexpected behavior of charm quarks at RHIC.

I stress, however, that RHIC 
is uniquely set to intensively study the region about $T_c$.
In the end, I feel no hestitation whatsoever in saying that
once RHIC turned on, we entered what is clearly a golden age in
high energy nuclear physics, one
which is well deserving of the highest possible recognition 
\cite{bjorken1,bjorken2,phenix,star}.

\end{document}